\documentclass{webofc}
\usepackage[varg]{txfonts}   
\begin{document}
\title{Gamma-widths, lifetimes and fluctuations \\in the
nuclear quasi-continuum}
%
%

\author{
\firstname{M.} \lastname{Guttormsen}\inst{1}\fnsep\thanks{\email{magne.guttormsen@fys.uio.no}} \and
\firstname{A.~C.} \lastname{Larsen}\inst{1}\fnsep\thanks{\email{a.c.larsen@fys.uio.no}} \and
\firstname{J.~E.} \lastname{Midtb{\o}}\inst{1} \and
\firstname{L.} \lastname{Crespo Campo}\inst{1} \and
\firstname{A.} \lastname{G{\"o}rgen}\inst{1} \and
\firstname{V.~W.~} \lastname{Ingeberg}\inst{1} \and
\firstname{T.} \lastname{Renstr{\o}m}\inst{1} \and
\firstname{S.} \lastname{Siem}\inst{1} \and
\firstname{G.~M.} \lastname{Tveten}\inst{1} \and
\firstname{F.} \lastname{Zeiser}\inst{1} \and
\firstname{L.~E.} \lastname{Kirsch}\inst{2}
}

\institute{Department of Physics, University of Oslo, N-0316 Oslo, Norway
\and
           Lawrence Berkeley National Laboratory, Berkeley, CA 94720, USA
}

\abstract{%
Statistical $\gamma$-decay from highly excited states is determined by the
nuclear level density (NLD) and the $\gamma$-ray strength function ($\gamma$SF).
These average quantities have been measured for several nuclei using the Oslo method.
For the first time, we exploit the NLD and $\gamma$SF to evaluate the
$\gamma$-width in the energy region below the neutron binding energy, often
called the quasi-continuum region. The lifetimes of states in the quasi-continuum
are important benchmarks for a theoretical description of nuclear structure and dynamics at
high temperature. The lifetimes may also have impact on reaction rates
for the rapid neutron-capture process, now demonstrated to take place in neutron star mergers.
}
\maketitle
\section{Introduction}
\label{intro}

Nature displays a huge span of lifetimes, from the birth and death
of stars to the population and decay of states in the micro-cosmos. In
the world of quantum physics, unstable states are associated with
an energy width $\Gamma$, which is related to the lifetime
through $\tau \Gamma=\hbar$. Both quantities depend on available
final states and the $\gamma$ strength into these states.

The nuclear level density (NLD) is an exponentially increasing function
of excitation energy. When the number of states reaches
100-1000 levels per MeV, detailed spectroscopy becomes
almost impossible and less useful. In this quasi-continuum region, the NLD
and the average $\gamma$-ray strength function ($\gamma$SF) become
fruitful concepts. These two quantities replace the accurate
position of initial and final states and the transition probabilities
between them in conventional discrete spectroscopy.

The Oslo method has provided NLDs and $\gamma$SFs
for many nuclei in the vicinity of the $\beta$-stability
line\footnote{Published data on NLDs and $\gamma$SFs
measured with the Oslo method are avaliable at {\it http://ocl.uio.no/compilation/}}.
From these observables, lifetimes, $\gamma$ widths, and fluctuations
can be explored in the quasi-continuum. In this work, we will
demonstrate the wealth of information that is hidden in these data.

The present study deals with the properties of levels in the quasi-continuum excitation region
below the neutron separation energy $S_n$. The level density is ranging from
around $10^3$ to $10^7$ levels per MeV at $S_n$, when going from nuclear masses of $A \sim 50$ to $240$.
For $\gamma$ energies around 3 MeV, the corresponding increase in $\gamma$ strength
is only one order of magnitude. This makes sense, because the NLD is fundamentally a combinatorial problem
of the number of active quasi-particles,
while the electric-dipole $\gamma$ strength scales linearly with the number of protons.

\section{The Oslo Method}
\label{sec:1}

In this section we give a short review of the Oslo method~\cite{Schiller00} for which the
starting point is a set of $\gamma$-ray spectra
measured as a function of initial excitation energy. The
$\gamma$ rays are measured in coincidence with the charged ejectile from light ion reactions
such as $(d, p\gamma)$, $(p,p'\gamma)$ and $(^3$He, $\alpha\gamma)$,
where the ejectile determines the initial excitation energy of each $\gamma$ spectrum.
Typical beam energies for the three reactions are 12 MeV, 16 MeV and 30 MeV, respectively.

\begin{figure}[]
\begin{center}
\includegraphics[width=\columnwidth]{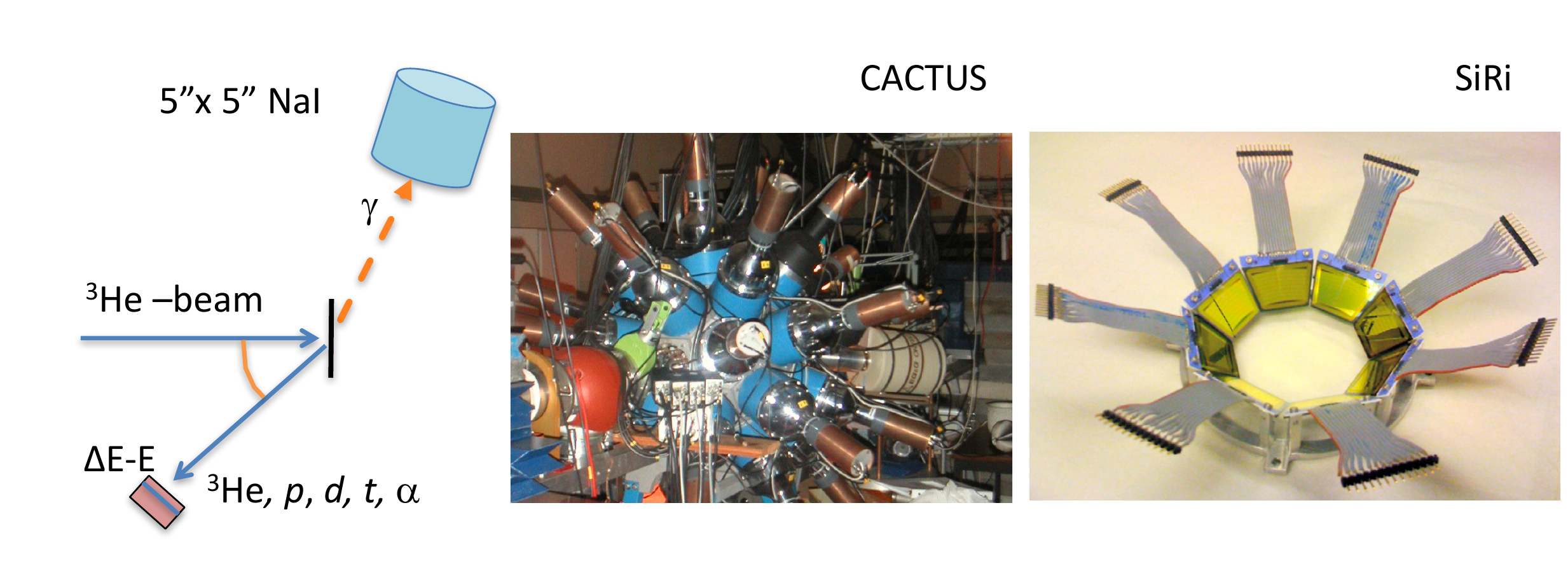}
\caption{(Color online) Typical particle-$\gamma$ coincidence set-up for the Oslo method.
The 64 silicon particle telescopes of SiRi are placed in the vacuum chamber at the center of CACTUS. }
\label{fig:set-up}
\end{center}
\end{figure}

Figure~\ref{fig:set-up} shows a schematic drawing of the set-up.
A silicon particle detection system (SiRi)~\cite{siri}, which
consists of 64 telescopes, is used for the selection of a certain ejectile types and to determine
their energies. The front $\Delta E$ and back $E$ detectors have thicknesses
of 130 $\mu$m and 1550 $\mu$m, respectively.
Coincidences with $\gamma$ rays are performed with the CACTUS array~\cite{CACTUS},
consisting of 26 collimated $5" \times 5"$ NaI(Tl) detectors with
a total efficiency of $14.1$\% at $E_\gamma = 1.33$~MeV.

With the raw $\gamma$-ray spectra at hand, we arrange these into a
particle-$\gamma$ matrix $R(E, E_{\gamma})$. Then, for all initial
excitation energies $E$, the $\gamma$ spectra are unfolded with the NaI response
functions giving the matrix $U(E, E_{\gamma})$~\cite{gutt1996}. The procedure is iterative and stops
when the folding ${\cal F}$ of the unfolded matrix equals the
raw matrix within the statistical fluctuations, i.e.~when ${\cal F}(U)\approx R$.

In the next step the primary $\gamma$-ray spectra are extracted from the unfolded matrix
$U$. This is obtained by subtracting a weighted sum of
$U(E',E_{\gamma})$ spectra below excitation energy $E$:
\begin{equation}
P(E,E_{\gamma})=U(E,E_{\gamma}) - \sum_{E' < E}w(E,E')U(E',E_{\gamma}).
\end{equation}
The weighting coefficients $w(E, E')$ are determined in an iterative way described in Ref.~\cite{Gut87}.
After a few iterations, $w(E,E')$ converges to $P(E,E_{\gamma})$, where we have normalized
each $\gamma$ spectrum by $\sum_{E_{\gamma}}P(E,E_{\gamma})=1$.
This conversion of $w \rightarrow P$ is exactly what is expected, namely that
the weighting function should equal the primary $\gamma$-ray spectrum.
We rely on the fact that quasi-continuum decay is dominated by dipole transitions~\cite{kopecky1990,larsen2013},
and consider only $E1$ and $M1$ transitions in the following.
It
should be noted that the validity of the
procedure rests on the assumption that the $\gamma$-energy
distribution is the same whether the levels were populated directly by
the nuclear reaction or by $\gamma$ decay from higher-lying states.

To extract the level density and the $\gamma$-ray strength function,
we exploit a part of the primary $P(E,E_{\gamma})$ matrix
where the level density is high, typically well above 2$\Delta$ (the pairing gap),
and where no single $\gamma$ lines dominate. This statistical part of the
matrix is described by the product of two vectors:
\begin{equation}
P(E, E_{\gamma}) \propto   \rho(E-E_{\gamma}){\cal{T}}(E_{\gamma}) ,\
\label{eqn:rhoT}
\end{equation}
where the decay probability is assumed to be proportional to the
NLD at the final energy $\rho(E-E_{\gamma})$ according
to Fermi's golden rule~\cite{dirac,fermi}. The decay is also proportional
to the $\gamma$-ray transmission coefficient ${\cal{T}}$, which
according to a generalized version of the Brink hypothesis~\cite{brink} is independent of spin and excitation energy;
only the transitional energy $E_{\gamma}$ plays a role.
The $\gamma$SF can be calculated from our measured transmission
coefficient through \cite{kopecky1990}
\begin{equation}
f_{XL} (E_{\gamma}) =\frac{1}{2 \pi}\frac{{\cal {T}}_{XL}(E_{\gamma})}{ E_{\gamma}^{2L+1}}.
\label{eq:f2t}
\end{equation}
It remains to normalize $\rho$ and ${\cal{T}}$ to known experimental information from other experiments.
The normalization procedures and the precisions obtained depend on available external data.
Further description and tests of the Oslo method and
the normalization procedures are given in Refs.~\cite{Schiller00,Lars11}.

One could argue that the level densities and $\gamma$SFs would depend on the light-ion
reaction used. However, although these reactions are very selective, the $\gamma$ decay appears
much later and thus from a thermalized, compound-like system.  This has been demonstrated by the Oslo group for
many reactions. As an example, the $(^3$He, $\alpha\gamma)$ and $(^3$He, $^3$He'$\gamma)$ reactions
have been studied populating the same final nuclei,
$^{96}$Mo and $^{97}$Mo~\cite{guttormsen2005,chankova2006}.
Also the very different reactions $(p,p'\gamma)$ and $(^3$He,
$\alpha\gamma)$ into $^{56}$Fe confirm the independence of
the reaction~\cite{larsen2013}. Minor differences may appear which
probably are due to deviations in the spin distributions populated by the various reactions.

\section{The evaluation of $\gamma$ width and lifetime}
\label{eval}

The $\gamma$ width ($\Gamma$) and lifetime ($\tau$) can be evaluated from the
measured NLD and $\gamma$SF obtained with the Oslo method. However, one should note
some differences when comparing with neutron capture data. First of all,
significantly more levels are populated in the charged-particle reaction, giving
a large spin distribution of typically $J \approx 0-6$ and populations of both parities.
Secondly, the initial excitation bin is much larger (100-200 keV) than
for neutron capture data that may even select only one ore a few resonances.
These conditions ensures that the Oslo type of data
represent an averaging over a broader initial excitation energy region and spins and parities.

The $\gamma$-decay strength function for $\gamma$-ray emission
of multipole $XL$ from levels of spin $J$ and parity $\pi$
at $E_x$ is defined by Bartholomew {\em et al.}~\cite{bart1973} as
\begin{equation}
f_{XL}^{J \pi}(E_{\gamma},E_x) = \frac{ \langle \Gamma_{XL}^{E_{\gamma}}(E_x,J,\pi)\rangle \rho(E_x,J,\pi)}{E_{\gamma}^{2L+1}},
\end{equation}
where $\langle \Gamma_{XL}^{E_{\gamma}}(E_x,J,\pi)\rangle$ is the {\em partial}
$\gamma$ width for the transition $E_x \rightarrow E_x-E_{\gamma}$. In the equation,
it is assumed that $E_x$ takes a fixed value while $E_{\gamma}$ takes
variable values, i.e. the final excitation energy
$E_x-E_{\gamma}$ varies. We now apply Eq.~(\ref{eq:f2t}) with the assumption
that $f_{XL}(E_{\gamma},E_x)$ is independent of $E_x$~\cite{brink}
and find
\begin{equation}
\langle \Gamma_{XL}^{E_{\gamma}}(E_x,J,\pi)\rangle=
\frac{1}{2\pi \rho(E_x,J,\pi)}{\cal T}_{XL}(E_{\gamma}).
\end{equation}
In order to obtain the the average {\em total} $\gamma$ width of levels with excitation energy $E_x $,
spin $J$ and parity $\pi$, we sum up the strength for all possible primary transitions below
$E_x$ as prescribed by Kopecky and Uhl~\cite{kopecky1990}:
\begin{equation}
\langle\Gamma(E_x,J,\pi)\rangle=\frac{1}{2\pi\rho(E_x, J, \pi)}
\sum_{J_f,\pi _f}\int_0^{E_x}{\mathrm{d}}E_{\gamma}{\cal T}_{XL}(E_{\gamma})
\times \rho(E_x-E_{\gamma}, J_f,{\pi}_f),
\label{eq:gipi}
\end{equation}
where the summation run over all final levels with spin $J_f$ and parity $\pi_f$
that are accessible by $XL$ transitions with energy $E_{\gamma}$. It should be noted
that for the normalization of ${\cal T}$, we apply Eq.~(\ref{eq:gipi}) using
the initial spin(s) populated in neutron capture to reproduce
the total $\gamma$ width $\langle \Gamma(S_n)\rangle$.
If we for simplicity assume that all levels within the initial energy bin
are populated in the charged particle reaction,
we obtain the average total $\gamma$ width by
\begin{equation}
\langle\Gamma(E_x)\rangle = \sum_{J\pi}g(E_x,J){\cal P}(E_x,\pi)\langle\Gamma(E_x,J,\pi)\rangle,
\label{eq:Gg}
\end{equation}
where $g$ and ${\cal P}$ are the spin and parity distributions, respectively.
From $\langle\Gamma(E_x)\rangle$, we get the
lifetime in the quasi-continuum by
\begin{equation}
\langle\tau(E_x)\rangle = \frac{\hbar}{\langle\Gamma(E_x)\rangle}= \frac{658.2 \ }{\langle\Gamma(E_x)\rangle}{\rm  fs},
\end{equation}
where the $\gamma$ width is in units of meV.
\begin{figure}[]
\begin{center}
\includegraphics[width=\columnwidth]{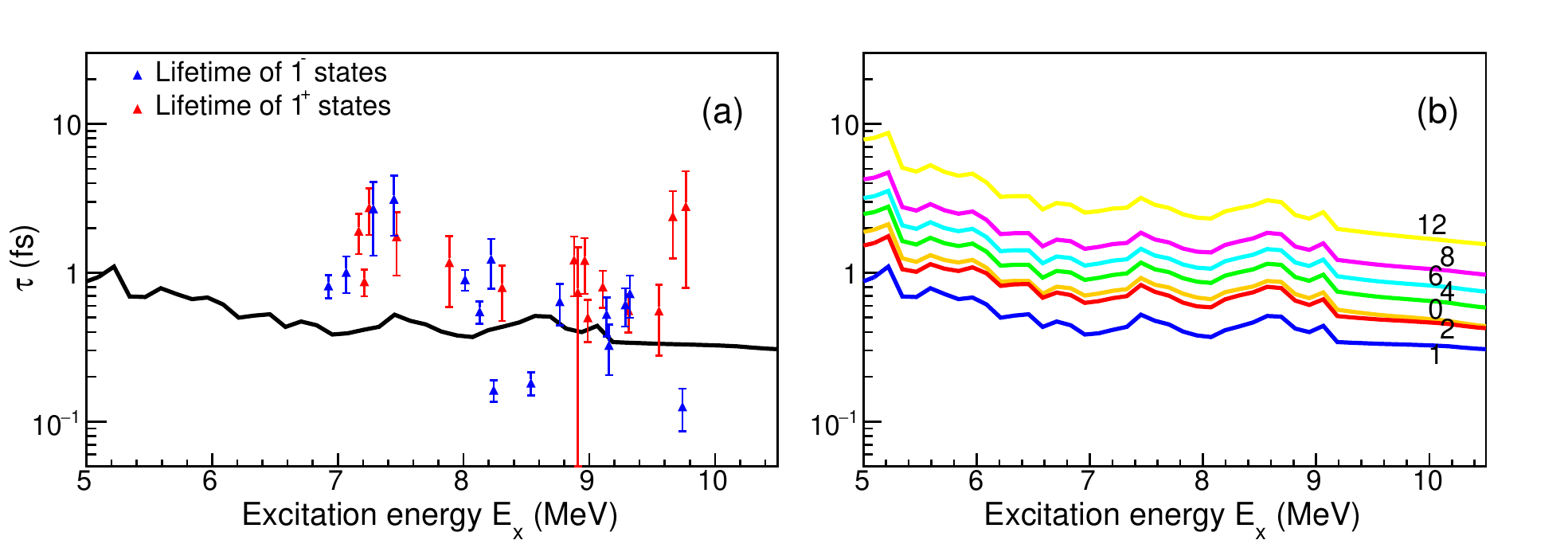}
\caption{(Color online) Experimental lifetimes of $J^{\pi} = 1^+$ and $1^-$
states in $^{56}$Fe~\cite{bauwens2000,shizuma2013} compared with
the estimates from the Oslo method (black line) (a).
The predictions for various spins are shown in panel (b).
}
\label{fig:life56fe}
\end{center}
\end{figure}

\begin{figure}[h]
\begin{center}
\includegraphics[width=0.8\columnwidth]{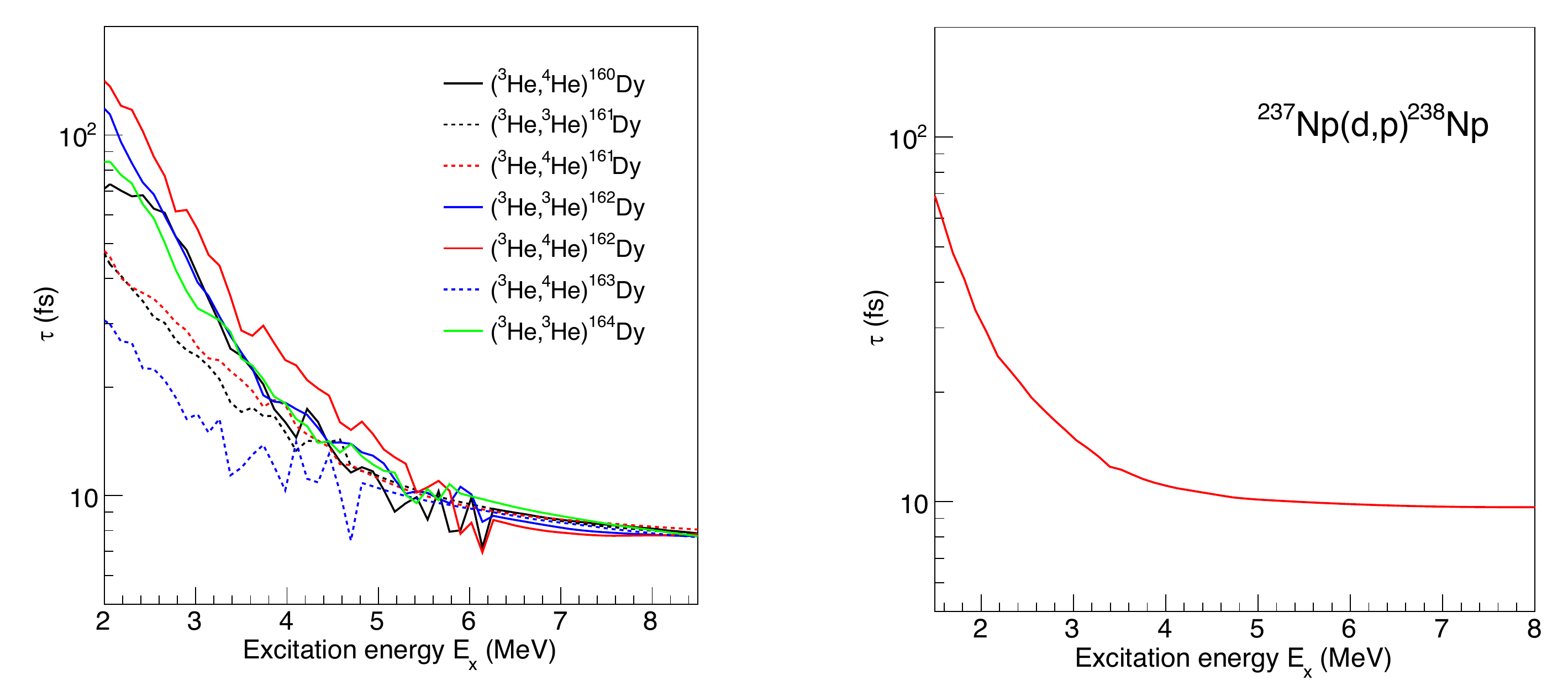}
\caption{(Color online) Lifetimes in the quasi-continuum region for $^{160-164}$Dy (a) and $^{238}$Np (b).
}
\label{fig:dynp}
\end{center}
\end{figure}

\section{Results and discussion}
\label{concl}

The $\gamma$ widths and lifetimes of $J^{\pi} = 1^+$ and $1^-$ states in the quasi-continuum of $^{56}$Fe have
been measured by photon-scattering experiments~\cite{bauwens2000,shizuma2013}. In Fig.~\ref{fig:life56fe} (a) we
show the measured lifetimes, and compare with the corresponding estimates of lifetimes
evaluated from the NLD and $\gamma$SF obtained from the Oslo method~\cite{larsen2013}.
We observe that the experimental data fluctuate up to a factor of ten,
which is a result of the random structural overlap with few final states.
Assuming Porter-Thomas fluctuations~\cite{PT}, the
relative fluctuations are $\sqrt{2/\nu}$ where the degree of freedom $\nu$
can be estimated by the number of primary transitions from the excited state. Figure~\ref{fig:life56fe} (b)
shows that the spin 1 states represent the fastest dipole transitions in the quasi-continuum,
which can be explained by their
direct decay to the $0^+$ ground and first excited $2^+$ states.

The number of levels in odd-mass dysprosiums are about a factor of seven larger than for the
even-even neighbours. This reduces the decay time for the odd-mass isotopes as seen in Fig.~\ref{fig:dynp} (a).
However, at higher excitations $(E_x \geq 5$ MeV) the lifetimes for all isotopes converge to a common value.
Furthermore, Fig.~\ref{fig:dynp} (b) demonstrates that the lifetimes of $^{238}$Np seem to flatten out
for $E_x \geq 4$ MeV. The saturation in $\tau$ seen for the two cases can be
understood from Eq.~(\ref{eq:gipi}): the decay by higher $\gamma$ energies for higher $E_x$
is suppressed by a factor $\exp{(-E_\gamma/T)}$ where $T$ is the nuclear temperature.
It is surprising that the saturation in lifetimes is the same ($8-10$ fs) for a large range of mass numbers.

\end{document}